\documentclass[12pt]{iopart}

\usepackage{iopams}
\usepackage{amssymb}
\usepackage{graphicx}
\eqnobysec

\begin{document}

\title[Geometric interpretation of the 3-dimensional coherence matrix]{Geometric interpretation of the 3-dimensional coherence matrix for nonparaxial polarization}

\author{M R Dennis}

\address{H H Wills Physics Laboratory, Tyndall Avenue, Bristol BS8 1TL, UK}

\begin{abstract}
The 3-dimensional coherence matrix is interpreted by emphasising its invariance with respect to spatial rotations.
Under these transformations, it naturally decomposes into a real symmetric positive definite matrix, interpreted as the moment of inertia of the ensemble (and the corresponding ellipsoid), and a real axial vector, corresponding to the mean angular momentum of the ensemble.
This vector and tensor are related by several inequalities, and the interpretation is compared to those in which unitary invariants of the coherence matrix are studied.
\end{abstract}


\section{Introduction}\label{sec:int}

In the standard theory of partial polarization in paraxial light \cite{fano:remarks,mw:quantum}, the $2\times 2$ hermitian coherence matrix (with unit trace) is decomposed into components with respect to the Pauli matrices. 
These components, the Stokes parameters, summarise the second order statistical information about the ensemble; in particular, the sum of their squares is 1 for a pure polarization state, and 0 for a completely unpolarized ensemble.

There has recently been a revival of interest in the corresponding coherence matrix in nonparaxial light, where in general there is no well-defined propagation direction, and the hermitian coherence matrix is $3\times 3$ \cite{samson:descriptions, barakat:degree, so:comments, brosseau:fundamentals, ckb:parameters, skf:thermal, sskf:degree}. 
In these treatments, by analogy with the 2-dimensional case, generalised Stokes parameters are defined by decomposing the coherence matrix with respect to the Gell-Mann matrices; a generalized degree of polarization \cite{samson:descriptions,barakat:degree,so:comments,skf:thermal,sskf:degree} may be defined using the sum of squares of these components.

Here, I propose a complementary interpretation of the $3\times 3$ coherence matrix, motivated by geometric reasoning. 
Rotational, rather than unitary, invariants of the coherence are emphasised, and the matrix is found to decompose into its real part, which is symmetric and interpreted geometrically as an ellipsoid, and its
imaginary part, which is antisymmetric and equivalent to an axial vector. 
The ellipsoid and vector have natural interpretation in terms of the ensemble of polarization states,
and are related by certain inequalities to be described.

Pure states are represented by a complex vector $\bi{E},$ representing the electric field, in either two or three dimensions. 
This is represented geometrically by an ellipse by taking $\rm{Re}\{ \bi{E} \exp(-\rmi \chi)\}$ and varying $\chi$ (this may represent time evolution) \cite{brosseau:fundamentals, nye:natural, dennis:polarization}; the ellipse therefore has a sense of rotation.
In two dimensions, this is taken in the natural sense with respect to the plane, and polarization is either right- or left-handed.
In three dimensions, the plane of the ellipse may vary, and the sense of rotation is a direction normal to the ellipse, defined in a right-handed sense with respect to the ellipse rotation \cite{nye:natural,bd:332}.
The eccentricity of the ellipse can be 1 (corresponding to linear polarization), 0 (corresponding to circular polarization), or any value in between.
The ellipses are normalised in units of intensity $|\bi{E}|^2.$
Polarization ensembles may be visualised geometrically as the set of polarization ellipses in the ensemble, adding incoherently.

The paper proceeds as follows: the following section is a review of conventional 2-dimensional coherence matrix theory; in section \ref{sec:geom}, the geometric decomposition of the
$3\times 3$ coherence matrix is described; section \ref{sec:ineq} is devoted to the properties of the coherence matrix, and section \ref{sec:exs} to examples for certain ensembles. 
The paper concludes with a discussion in section \ref{sec:disc}.

Polarization coherence matrices are special (classical) occurrences of density matrices, perhaps more familiar in quantum mechanics \cite{fano:description,sakurai:modern} (pure polarization states corresponding to pure states, etc). 
Standard properties of density matrices (i.e. positive definite matrices with unit trace) will be employed without proof.

\section{The two-dimensional coherence matrix}\label{sec:2d}

This section is included as a comparison for the $3\times 3$ case, and reviews standard material discussed, for example, in \cite{fano:remarks,mw:quantum,brosseau:fundamentals}.

The 2-dimensional coherence matrix $\rho_2,$ assumed normalised (i.e. $\tr \rho_2 = 1$), is defined
\begin{equation}
   \rho_2  = \left( \begin{array}{cc} \langle E_x  E_x^{\ast} \rangle & \langle E_x E_y^{\ast}  \rangle \\
\langle E_y E_x^{\ast} \rangle & \langle E_y E_y^{\ast} \rangle
\end{array} \right),
   \label{eq:2ddef}
\end{equation}
where $\langle \bullet \rangle$ denotes ensemble averaging over the ensemble of 2-dimensional complex vectors $\bi{E} = (E_x, E_y).$ 
$\rho_2$ is normally expressed in terms of the \emph{Stokes parameters} $S_1, S_2, S_3,$ which are the components of $\rho_2$ with respect to the Pauli matrices:
\begin{equation}
   \rho_2 = \frac{1}{2} \left( \begin{array}{cc} 1 + S_1 & S_2 - \rmi S_3 \\
S_2 + \rmi S_3 & 1 - S_1
\end{array} \right).
   \label{eq:2dpauli}
\end{equation}
The three Stokes parameters may be written as a 3-vector, the
\emph{Stokes vector}
\begin{equation}
   \bi{P} = (S_1, S_2, S_3)
   \label{eq:stokesv}
\end{equation}
whose length $|\bi{P}|$ is written $P.$

Being a density matrix, $\rho_2$ is positive definite (its eigenvalues are nonnegative), so
\begin{equation}
   \det \rho_2 = (1- S_1^2 - S_2^2 - S_3^2)/4 \ge 0,
   \label{eq:detrho2}
\end{equation}
that is,
\begin{equation}
   P \le 1,
  \label{eq:pineq}
\end{equation}
which geometrically restricts $\bi{P}$ to lie within a sphere of radius 1, the \emph{Poincar{\'e} sphere.} 
This fundamental inequality is more commonly derived using the equivalent fact $\tr \rho_2^2 \le (\tr \rho_2)^2.$

If the ensemble represents a single state of polarization (i.e. a `pure state'), the coherence matrix is idempotent,
\begin{equation}
   \rho_{2,\rm{pure}}^2 = \rho_{2,\rm{pure}}.
   \label{eq:idempotent}
\end{equation}
Taking the trace implies that $P_{\rm{pure}} = 1.$ 
On the other hand, if the ensemble is completely unpolarized, so $\rho_{2,\rm{un}}$ is $1/2$ times the identity matrix, then $P_{\rm{un}} = 0.$ 
This leads to the important decomposition of $\rho_2$ into pure and unpolarized parts,
\begin{equation}
   \rho_2 = (1-P) \rho_{2,\rm{un}} + P \rho_{2,\rm{pure}}.
   \label{eq:2ddecomp}
\end{equation}
The state of polarization corresponding to $\rho_{2,\mathrm{pure}}$ here is the eigenvector corresponding to the larger eigenvalue of $\rho_2,$ and $1 - P$ is twice the
smaller eigenvalue.

The previous statements justify $P$ as the \emph{degree of polarization}. 
It is invariant with respect to any unitary transformation $\mathbf{u} \rho_2 \mathbf{u}^{\dagger},$ by (\ref{eq:detrho2}) (here, $\mathbf{u}$ represents an arbitrary
$2\times 2$ unitary matrix). 
By the well-known relation between $2\times 2$ unitary and $3\times 3$ orthogonal matrices, such unitary transformations correspond to rotations of the Stokes vector $\bi{P}.$ 
The operation of a unitary transformation on polarization states (or their ensemble average) is physically interpreted as the operation of a phase retarder \cite{brosseau:fundamentals}, and the degree of polarization is unchanged when the ensemble is passed through a retarder, or series of them.

The Stokes vector (\ref{eq:stokesv}) resides in an abstract, 3-dimensional (Stokes) space, and the representation of phase retarders by 3-dimensional rotations is correspondingly abstract.
If $\rho_2$ is transformed by 2-dimensional rotations, corresponding to a real rotation of the transverse plane (i.e. $\mathbf{o} \rho_2 \mathbf{o}^{\mathrm{T}},$ with $\mathbf{o}$ $2\times 2$ orthogonal), $S_1$ and $S_2$ may change keeping $S_1^2 + S_2^2$ constant; $S_3$ remains unchanged. 
An example case is the rotation in which $\rm{Re} \rho_2$ is diagonalised:
\begin{equation}
   \rho_{2,\mathrm{rot}} = \frac{1}{2} \left( \begin{array}{cc} 1 + \sqrt{S_1^2 +S_2^2} &  - \rmi S_3 \\
\rmi S_3 & 1-\sqrt{S_1^2 +S_2^2}
\end{array} \right).
   \label{eq:rho2rot}
\end{equation}

For pure states, for which the Poincar{\'e} sphere representation is useful, the Stokes parameters provide geometric information about the polarization ellipse \cite{brosseau:fundamentals,dennis:polarization}. 
$S_1$ and $S_2$ inform about the alignment of the ellipse axes, the major axis making an angle $\arg(S_1 + \rmi S_2)/2$ with the $x$-axis. 
$S_3$ gives the ellipse area $\pi S_3,$ signed according to polarization handedness, so
$S_3$ is zero for linear, and $\pm1$ for circular polarization. 
Obviously, 2-dimensional rotations only affect $S_1$ and $S_2;$ the rotation giving (\ref{eq:rho2rot}) represents aligning the major ellipse axis along $x,$ the minor along $y.$

\section{Geometry of the 3-dimensional coherence matrix}\label{sec:geom}

The 3-dimensional coherence matrix $\rho = \rho_3$ is analogous to
(\ref{eq:2ddef}), but with $\bi{E} = (E_x, E_y, E_z):$
\begin{equation}
   \rho = \left( \begin{array}{ccc} \langle E_x  E_x^{\ast} \rangle & \langle E_x E_y^{\ast} \rangle &  \langle E_x E_z^{\ast}  \rangle \\
\langle E_y E_x^{\ast} \rangle & \langle E_y E_y^{\ast} \rangle & \langle E_y E_z^{\ast} \rangle \\
\langle E_z E_x^{\ast} \rangle & \langle E_z E_y^{\ast} \rangle & \langle E_z E_z^{\ast} \rangle
\end{array} \right).
   \label{eq:3ddef}
\end{equation}
As before, it is assumed that $\tr \rho = 1.$

The Gell-Mann matrices \cite{griffiths:introduction}  are the generators of 3-dimensional unitary matrices, just as the Pauli matrices generate 2-dimensional unitary matrices.
Therefore, the \emph{generalised Stokes parameters} $\Lambda_i,$ $i=1,\dots, 8$ \cite{barakat:degree, brosseau:fundamentals, ckb:parameters, skf:thermal, sskf:degree},                                                                                                                                may be defined 
\begin{equation}
   \rho = \frac{1}{3} \left( \begin{array}{ccc}
1+\Lambda_3+\Lambda_8/\sqrt{3} & \Lambda_1 - \rmi \Lambda_2 & \Lambda_4 - \rmi \Lambda_5 \\
\Lambda_1 + \rmi \Lambda_2 & 1-\Lambda_3+\Lambda_8/\sqrt{3} & \Lambda_6 - \rmi \Lambda_7 \\
\Lambda_4 + \rmi \Lambda_5 & \Lambda_6 + \rmi \Lambda_7 & 1- 2 \Lambda_8/\sqrt{3} \end{array} \right) .
   \label{eq:gellman}
\end{equation}
(Other accounts, such as \cite{samson:descriptions}, use a different set of generators.) The analogies between (\ref{eq:2dpauli}) and (\ref{eq:gellman}) are obvious: $\Lambda_3$ and $\Lambda_8,$ only appearing on the diagonal, generalise $S_1;$ the terms in the symmetric, off-diagonal part, $\Lambda_1, \Lambda_4, \Lambda_7,$ generalise $S_2;$
and $\Lambda_2, \Lambda_5, \Lambda_7,$ appearing in the antisymmetric, imaginary part, $S_3.$ 
In particular, if $\Lambda_4, \dots, \Lambda_7 = 0$ and $\Lambda_8 = \sqrt{3}/2,$ then the remaining parameters are proportional to the usual Stokes parameters. 
This motivates the definition of the \emph{generalised degree of polarization} $P_3$ \cite{barakat:degree, skf:thermal, sskf:degree} as
\begin{equation}
   P_3 = \sqrt{\sum_{i=1}^{8} \Lambda_i^2/3}.
   \label{eq:3ddeg}
\end{equation}
(A slightly different form was defined by \cite{samson:descriptions,so:comments}.) This is the length of the 8-dimensional generalised Stokes vector, and, just as in the 2-dimensional case, it is invariant with respect to $3\times 3$ unitary transformations. Since $(\tr \rho)^2 - \tr\rho^2 \ge 0,$ it is readily shown that $0 \le P_3 \le 1.$

Although this approach is mathematically correct, it is not clear physically what $P_3$ represents.
Unlike the $2\times 2$ case, in which the Stokes vector represents the complete state of
polarization using three dimensions (which is easily visualised), the generalised Stokes vector requires eight dimensions, which is not so intuitive.

There is a more serious problem with treating the $3\times 3$ coherence matrix completely in analogy with the $2 \times 2$ case - there is no obvious physical interpretation via optical elements of $3 \times 3$ unitary transformations (nor any corresponding nonparaxial Jones or Mueller calculus). 
In two dimensions, as an ensemble of plane waves with the same direction but different polarizations  propagates through an optical element, the corresponding coherence matrix is transformed by the appropriate Jones matrix, which is unitary for a retarder.
In three dimensions, the ensemble of plane waves averaging to the $3\times 3$ coherence matrix do not share a common propagation direction in general; any physical device,  represented by a $3\times 3$ unitary transformation, should be insensitive to the propagation directions of the separate members of the ensemble. 
Mathematically, it is possible to find a unitary transformation which takes any 3-dimensional state of polarization $\bi{E} = (E_x,E_y,E_z)$ to any other (leaving $|\bi{E}|^2$ constant); there is no obvious physical situation in which different states of polarization in three dimensions undergo the same unitary transformation.

It is physically and geometrically natural, however, to consider $\rho_3$ under orthogonal transformations rather than unitary ones; if viewed as passive rotations, this is simply equivalent to redefining cartesian axes in 3-dimensional space, and no
physical operation at all. 
Clearly, under rotation, where $\rho$ becomes $\mathbf{o} \rho \mathbf{o}^{\mathrm{T}}$ ($\mathbf{o}$ $3\times 3$ orthogonal), the real and imaginary parts of $\rho$ transform independently of each other. 
The real part is a positive definite symmetric matrix with five parameters $\Lambda_1, \Lambda_3, \Lambda_4, \Lambda_6, \Lambda_8.$ 
Since the (unit) trace is also unaffected by rotation, it may be considered as distinct from the rest of the real part. 
The imaginary part is a real antisymmetric matrix with three parameters $\Lambda_2,
\Lambda_5, \Lambda_7,$ and in fact the triple $(\Lambda_7, -\Lambda_5, \Lambda_2)$ transforms under rotation like an axial vector (noted in \cite{ckb:parameters}). 
$\rho$ therefore decomposes into three parts: a real scalar (the trace), a real axial vector, and a real traceless symmetric matrix. 
These different parts (scalar, vector, tensor) are called \emph{irreducible tensor operators} in group theory; the same decomposition occurs for density matrices of atoms with quantum spin 1, for which the vector part is called the \emph{orientation}, the tensor part the \emph{alignment} \cite{blum:density}.

From an analytical viewpoint, it is convenient to represent $\rho$ using cartesian axes $x_1, x_2, x_3$ with respect to which the tensor part is diagonal, giving
\begin{equation}
   \rho = \left(\begin{array}{ccc}
M_1 & -\rmi N_3 & \rmi N_2 \\
\rmi N_3 & M_2 & -\rmi N_1 \\
-\rmi N_2 & \rmi N_1 & M_3 \end{array} \right).
   \label{eq:rho123}
\end{equation}
The diagonal elements of (\ref{eq:rho123}) are restricted:
\begin{equation}
   M_1 + M_2 + M_3 = 1, \qquad 1 \ge M_1 \ge M_2 \ge M_3 \ge 0,
   \label{eq:mrest}
\end{equation}
which follows from the fact that the tensor $\mathbf{M} \equiv \mathrm{Re} \rho$ is positive definite. 
It is geometrically convenient not to separate the scalar and (traceless) tensor parts of $\rho,$ and this is not done in (\ref{eq:rho123}).
(\ref{eq:rho123}) is analogous to (\ref{eq:rho2rot}); the real part $\mathbf{M}$ has been (passively) diagonalised, leaving an off-diagonal imaginary part, which transforms as an axial vector
\begin{equation}
   \bi{N} = (N_1, N_2, N_3)
   \label{eq:ndef}
\end{equation}
($|\bi{N}|$ is invariant under rotations).
$\mathbf{M}$ and $\bi{N}$ have a simple geometrical interpretation, as follows.

The real symmetric matrix $\mathbf{M}$ may be interpreted as the moment of inertia tensor of the ensemble. 
Geometrically, it is the moment of inertia of the set of polarization ellipses in the
ensemble (taking each as an elliptical ring with uniform mass per unit length, insensitive to the ellipse handedness). 
As with moment of inertia tensors in mechanics, it may be represented in terms of its \emph{inertia ellipsoid}, whose points $(x_1, x_2, x_3)$ satisfy
\begin{equation}
   \frac{x_1^2}{M_1} + \frac{x_2^2}{M_2} + \frac{x_3^2}{M_3} = 1.
   \label{eq:moiellipsoid}
\end{equation}
The ellipsoid axes are aligned in the 1,2,3 directions, with lengths $\sqrt{M_1}, \sqrt{M_2}, \sqrt{M_3}.$ 
If $M_3 = 0,$ the ellipsoid is flat ($x_3 = 0$). 
In general, the inertia ellipsoid is specified by 6 parameters (the trace and $\Lambda_1, \Lambda_3, \Lambda_4, \Lambda_6, \Lambda_8$); the diagonal form in (\ref{eq:rho123}), with three parameters, reflects that three Euler angles have been used implicitly in the choice of axes 1,2,3. 
The traceless part, dependent on the $\Lambda$ parameters only, gives a measure of departure of this inertia tensor from isotropy.

The vector $\bi{N}$ also has a simple interpretation as half the expectation value for (spin) angular momentum in the ensemble,
\begin{equation}
   \overline{\bi{S}} = \tr(\mathbf{S} \rho) = 2 \bi{N},
   \label{eq:spinexp}
\end{equation}
where the spin matrices $\mathbf{S}_i$ for spin 1 in a cartesian basis are given componentwise by $S_{i,jk} = -\rmi \varepsilon_{ijk}$ \cite{altmann:rotations, bd:332}, with $\varepsilon_{ijk}$ the antisymmetric symbol. 
The axial vector $\bi{N}$ is therefore an average of the angular momentum, that is, the average sense of rotation of the ellipses, in the ensemble.
Its direction, in general, has no relation to the principal axes of $\mathbf{M}$ (although its maximum length is limited by them, as described in the next section). 

The inertia tensor $\mathbf{M}$ and orientation vector $\bi{N}$ therefore provide information about the real, 3-dimensional geometry of the polarization ensemble, and they rotate rigidly.
Under more general unitary transformations (which have no physical interpretation), the eigenvalues of $\mathbf{M}$ and components of $\bi{N}$ may change arbitrarily (although keeping the unitary invariants $\tr \rho, \tr \rho^2$ and $\det \rho$ fixed).

\begin{figure}
\begin{center}
\includegraphics*[width=10cm]{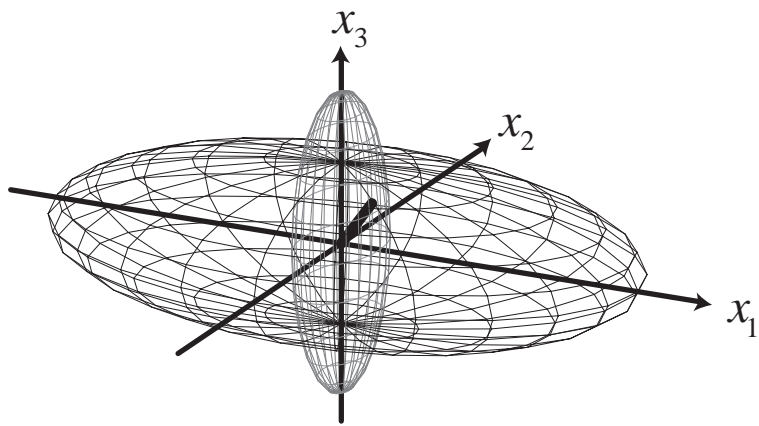}
\end{center}
    \caption{The inertia ellipsoid (black mesh), dual ellipsoid (grey mesh) and orientation vector corresponding to $\rho_{\rm{ex}},$ in the $x_1, x_2, x_3$ frame. Here, the orientation vector lies inside the dual ellipsoid, and not on its surface.}
    \label{fig:ex}
\end{figure}

As an example of the geometric interpretation, figure \ref{fig:ex} is a representation of the inertia ellipsoid, orientation vector and dual ellipsoid (defined in the next section) for the matrix
\begin{equation}
   \rho_{\rm{ex}} = \frac{1}{20} \left( \begin{array}{ccc} 14 & -2\rmi  & 2\rmi  \\ 2\rmi  & 5 & -\rmi  \\ -2 \rmi  & \rmi  & 1 \end{array}\right).
   \label{eq:ex}
\end{equation}

$\rho$ may also be represented by its eigenvectors; if $\rho_a, \rho_b, \rho_c$ represent the pure, idempotent coherence matrices corresponding to the eigenvectors of $\rho$ with eigenvalues $\lambda_a, \lambda_b, \lambda_c$ (i.e. the principal idempotents \cite{samson:descriptions,barakat:degree}), then 
\begin{equation}
   \rho = \lambda_a \rho_a + \lambda_b \rho_b  + \lambda_c \rho_c.
   \label{eq:3ddecomp}
\end{equation}
The 2-dimensional analogue to (\ref{eq:3ddecomp}) immediately gives rise to the decomposition (\ref{eq:2ddecomp}).
Since the decomposition in (\ref{eq:3ddecomp}) is in terms of three density matrices, a decomposition in terms of a single purely polarized part and unpolarized part is, in general, impossible (as previously noted in \cite{barakat:degree, sskf:degree}). 
The eigenvalues of $\rho,$ being unitary invariants, do not have a simple geometric interpretation in terms of the inertia ellipsoid or orientation vector; however, the set of three eigenvectors rotates rigidly.
This eigenvector representation of $\rho$ provides a different geometric representation to that given by the inertia ellipsoid and orientation vector.
However, it not not geometrically obvious when a given triple of polarization ellipses in three dimensions represent orthogonal polarization states, and solution of cubic equations is required to find the eigenvectors; moreover, the eigenvectors are not uniquely defined at a degeneracy.
By comparison, the inertia tensor and orientation vector may be extracted directly from $\rho$ and are always unambiguously defined.

\section{Inequalities satisfied by $\rho$}\label{sec:ineq}

In this section, various inequalities for $\mathbf{M}$ and $\bi{N}$ shall be found, using the fact that $\rho$ is a statistical density matrix.

Firstly, the Cauchy-Schwartz inequality may be applied to the off-diagonal elements of $\rho$ in (\ref{eq:3ddef}), giving expressions of the form
\begin{equation}
   |\langle E_x E_y^{\ast} \rangle |^2 \le \langle | E_x |^2 \rangle \langle | E_y |^2 \rangle.
   \label{eq:csdef}
\end{equation}
Using the representation (\ref{eq:rho123}), these imply
\begin{equation}
   N_1^2 \le M_2 M_3,\qquad  N_2^2 \le M_1 M_3, \qquad N_3^2 \le M_1 M_2.
   \label{eq:cs}
\end{equation}
Geometrically, this implies that the orientation vector $\bi{N}$ is confined to a cuboid with vertices $(\pm \sqrt{M_2 M_3}, \pm \sqrt{M_1 M_3}, \pm \sqrt{M_1 M_2}).$ 
Since $\rho$ is a density matrix, $(\tr \rho)^2 - \tr \rho^2 \ge 0,$ that is
\begin{equation}
   N_1^2 + N_2^2 + N_3^2 \le M_2 M_3 + M_1 M_3 + M_1 M_2,
   \label{eq:trsquare}
\end{equation}
which is the sum of the inequalities (\ref{eq:cs}), and therefore is less strong, geometrically restricting $\bi{N}$ to lie within the sphere circumscribing the cuboid defined above. 
$(\tr \rho)^2 - \tr \rho^2 \ge 0,$ which is the distance by which $\bi{N}$ fails to touch the surface of this sphere, is a unitary invariant. 
The traces of higher powers of $\rho$ satisfy other inequalities, such as $\tr
\rho^3 \le \tr \rho^2 \tr \rho,$ but such inequalities can be shown
to be consequences of (\ref{eq:cs}).

Nonnegativity of $\det \rho$ implies that
\begin{equation}
   M_1 N_1^2 + M_2 N_2^2 + M_3 N_3^2 \le M_1 M_2 M_3.
   \label{eq:det}
\end{equation}
If $M_3 \neq 0,$ then
\begin{equation}
   \frac{N_1^2}{M_2 M_3} + \frac{N_2^2}{M_1 M_3} + \frac{N_3^2}{M_1 M_2} \le 1,
   \label{eq:dualellipsoid}
\end{equation}
which geometrically means that $\bi{N}$ lies within the ellipsoid with axes in the 1,2,3 directions, and lengths $\sqrt{M_2 M_3}, \sqrt{M_1 M_3}, \sqrt{M_1 M_2}.$ 
This ellipsoid is therefore circumscribed by the cuboid (\ref{eq:cs}), and (\ref{eq:dualellipsoid}) is a stronger inequality than (\ref{eq:cs}). 
The relationship between this ellipsoid and the inertia ellipsoid (\ref{eq:moiellipsoid}) justifies calling this ellipsoid  the \emph{dual ellipsoid}. 
If $M_3 = 0,$ (\ref{eq:det}) implies that $N_1 = N_2 = 0,$ and (\ref{eq:cs}) gives $|N_3| \le \sqrt{M_1 M_2};$ if the inertia ellipsoid is flat, the dual ellipsoid is a line normal to it. 
If $M_1 = 1, M_2 = M_3 =0,$ then the inertia ellipsoid is a line and $\bi{N} = 0.$

As with (\ref{eq:detrho2}), the fundamental inequality for the $3 \times 3$ coherence matrix is nonnegativity of the determinant, which is stronger than inequalities constructed using the trace.
The geometric interpretation of the unitary invariant $\det \rho$ is the product of the distance by which $\bi{N}$ fails to touch the dual ellipsoid with the dual ellipsoid volume. This quantity, the trace, and the invariant discussed above are the only unitary invariants of $\rho.$ 
Unlike the 2-dimensional case, the properties of $\rho$ are complicated by the fact that polarization information is contained within both the inertia ellipsoid $\mathbf{M}$ and the orientation vector $\bi{N}.$

\section{Examples of $3\times 3$ polarization ensembles}\label{sec:exs}

Completely unpolarized waves in three dimensions are a common occurrence, for example black body radiation. In this situation, the $3\times 3$ coherence matrix is the completely unpolarized matrix $\rho_{\rm{un}},$ equal to one third times the $3\times 3$ identity matrix (and $P_3 = 0$). 

Coherence matrices for pure states of polarization satisfy $\rho _{\mathrm{pure}}^2 = \rho_{\mathrm{pure}}.$ 
Using (\ref{eq:rho123}) and (\ref{eq:mrest}), this implies that
\begin{equation}
   \rho_{\mathrm{pure}} = \left( \begin{array}{ccc} M_1 & -\rmi N_3 & 0 \\ \rmi N_3 & M_2 & 0 \\ 0 & 0 & 0 \end{array} \right)
   \label{eq:rhopure}
\end{equation}
with $|N_3| = \sqrt{M_1 M_2}, M_1 + M_2 = 1.$ 
This is equivalent to a pure state in two dimensions, and represents a polarization
ellipse $\bi{E} = (\sqrt{M_1}, \pm\rmi \sqrt{M_2}, 0)$ in 1,2,3 coordinates. 
The ellipse major axis is in the 1-direction, the minor in the 2-direction, and $\bi{N}$ is
normal to the plane of the ellipse (oriented in a right-handed sense of rotation around the ellipse). 
If $M_1 = M_2 = 1/2$ in (\ref{eq:rhopure}), the state is circularly polarized, and $N_3 =
\pm 1/2.$ 
If $M_1 = 1, M_2 = 0$ (implying $N_3 = 0$), it is linearly polarized.
$\det \rho_{\mathrm{pure}}$ is zero, but unlike the $2\times 2$ case this is not a sufficient condition for a pure state in general: $\tr \rho^2$ must also be 1. 
The inertia ellipsoid of (\ref{eq:rhopure}) is flat, and $\bi{N}$ lies on the `surface' of the (linear) dual ellipsoid, with equality in (\ref{eq:cs}).

If the state is not pure but $M_3 = 0,$ then $\rho$ satisfies (\ref{eq:rhopure}) with $M_1 + M_2 =1,$ but $|N_3| < \sqrt{M_1 M_2}.$ 
An example is the density matrix
\begin{equation}
   \rho_{\mathrm{ex1}} = \frac{1}{3} \left(\begin{array}{ccc} 2 & 0 & 0 \\ 0 & 1 & 0 \\ 0 & 0 & 0 \end{array} \right).
   \label{eq:ex1}
\end{equation}
The inertia ellipsoid here is flat, and $\bi{N} = 0.$ 
It cannot be a pure state since $\rho_{\mathrm{ex1}}^2 \neq \rho_{\mathrm{ex1}}.$ 
This matrix provides an example of a $3\times 3$ coherence matrix which cannot be decomposed into the sum of a pure polarization matrix and the completely unpolarized matrix, since there is a zero on the diagonal - $\rho_{\mathrm{ex1}} - \alpha \rho_{\mathrm{un}},$ for any positive $\alpha,$ leaves a matrix which is not positive definite.

It is easy to visualise ensembles which have $\bi{N} = 0$: their average angular momentum  is zero. 
This may be achieved, for instance, by requiring for every $\bi{E}$ in the ensemble, $\bi{E}^{\ast}$ has the same statistical weight as $\bi{E}.$
$\rho_{\rm{ex1}}$ is therefore the coherence matrix for the ensemble consisting of the pair of states (with equal weight)
\begin{equation}
   \mathcal{E}_{\mathrm{ex1}}  = \{ (\sqrt{2}, \rmi,0), (\sqrt{2}, -\rmi, 0)\}
   \label{eq:ensex1}
\end{equation}
(of course, this ensemble is not unique in averaging to $\rho_{\mathrm{ex1}}$). 
The ellipses corresponding to the pair (\ref{eq:ensex1}) are identical apart from their senses of rotation, which are opposite.

\begin{figure}
\begin{center}
\includegraphics*[width=8cm]{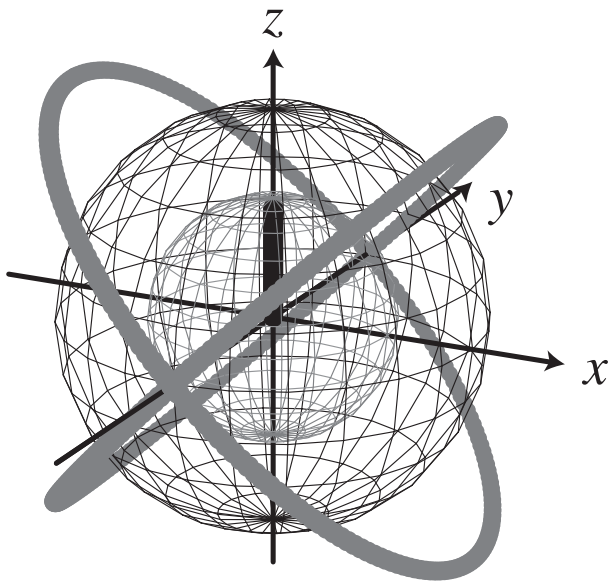}
\end{center}
    \caption{The pair of polarization ellipses corresponding to the ensemble (\ref{eq:ensex2}) (grey), with their spherical inertia ellipsoid (black mesh), spherical dual ellipsoid (grey mesh) and orientation vector, which here is vertical and on the surface of the dual ellipsoid.}
    \label{fig:ex2}
\end{figure}

$\rho_{\mathrm{ex1}}$ is an example of a coherence matrix with $\bi{N} = 0,$ although its $\mathbf{M}$ is not isotropic; that is, the shape of the inertia ellipsoid is not constrained by the direction of the orientation vector. 
More surprising, perhaps, is that the converse is true - the inertia ellipse may be isotropic
yet $\bi{N}$ takes on the maximum value allowed by (\ref{eq:cs}), for example
\begin{equation}
   \rho_{\mathrm{ex2}} = \frac{1}{3} \left( \begin{array}{ccc} 1 & -\rmi & 0 \\ \rmi & 1 & 0 \\ 0 & 0 & 1\end{array}\right),
   \label{eq:ex2}
\end{equation}
which is the sum of $\rho_{\mathrm{un}}$ and a completely antisymmetric matrix (which is not a density matrix). 
An ensemble which corresponds to $\rho_{\mathrm{ex2}}$ is the pair of states with equal weight
\begin{equation}
   \mathcal{E}_{\mathrm{ex2}} = \{(1,\rmi,-1), (1,\rmi,1)\}.
   \label{eq:ensex2}
\end{equation}
The ellipses represented here share their minor axis (in the  $y$-direction) and have orthogonal major axes. 
They both have the same shape (eccentricity $1/\sqrt{2}$), which geometrically implies
that their total moment of inertia is isotropic (higher averages than quadratic are not isotropic).
This pair of ellipses, along with the spherical inertia ellipsoid and orientation vector, are shown in figure \ref{fig:ex2}.

Both $\rho_{\mathrm{ex1}}$ and $\rho_{\mathrm{ex2}}$ have the same eigenvalues $2/3, 1/3, 0$ (equivalently, the same unitary invariants $\tr \rho, \tr \rho^2, \det \rho$); however, the two ensembles (\ref{eq:ensex1}), (\ref{eq:ensex2}) are clearly not the same: the ellipses in the two ensembles have the same shape (eccentricity $1/\sqrt{2}$), but the orientations in space are different, and there is no obvious physical transformation between the two sets of states.

In general, the minimum number of states in an ensemble required to specify $\rho$ is three, and in fact, the (complex) eigenvectors of $\rho$ suffice, as in (\ref{eq:3ddecomp}). 
In this case, the eigenvectors make up the ensemble, the probability weighting for each being the corresponding eigenvalue.
Since $\rho_{\mathrm{ex1}}, \rho_{\mathrm{ex2}}$ each have one zero eigenvalue, an ensemble consisting only of two states is sufficient for these examples (the states in $\mathcal{E}_{\mathrm{ex1}}, \mathcal{E}_{\mathrm{ex2}}$ are linear combinations of the eigenvectors, and are not orthogonal).

\section{Discussion}\label{sec:disc}

Interfering nonparaxial polarization fields in three dimensions are more complicated than their paraxial counterparts, and their analysis involves subtle geometric reasoning \cite{nye:natural,
dennis:thesis, bd:332,dennis:polarization}. 
Most importantly, the Poincar{\'e} sphere description breaks down for polarization states in three dimensions, because it cannot account for the direction of the ellipse normal $\bi{N};$ the appropriate nonparaxial analogue of the Poincar{\'e} sphere is the \emph{Majorana sphere}, which involves the symmetric product of two unit vectors, which describe the geometry of the nonparaxial
polarization ellipse \cite{penrose:emperors, hannay:majorana,
dennis:thesis}. 
These two vectors have a complicated expression in terms of the pure field state $\bi{E}.$

It would be of interest to find the relationship between the $3 \times 3$ coherence matrix and ensembles defined in terms of the Majorana sphere; a natural physical case would be when the $E_x, E_y, E_z$ field components are gaussian distributed (for example black body radiation). 
In this case, for a given $\rho$ the distribution on the Majorana sphere would be unique and related to other gaussian Majorana statistics \cite{hannay:zero}. 
The analogous $2 \times 2$ distributions on the surface of the Poincar{\'e} sphere have a rather simple form \cite{brosseau:fundamentals,barakat:stokes,eliyahu:stokes,brosseau:statistics}. 
Given the analytical complications of the Majorana sphere, it is unlikely that the $3 \times 3$ calculations will be straightforward, and it is unclear whether the geometric interpretation presented here would be helpful in this problem.

As described in section \ref{sec:geom}, there is no unique direction, or set of directions, associated with propagation for a general $3 \times 3$ $\rho,$ and therefore, unlike the $2\times 2$ case, there is no physical interpretation of $3\times 3$ unitary transformations using conventional optical elements. 
Despite this lack of propagation information, a natural application of the 3-dimensional coherence matrix is in scattering theory, since a scatterer, such as a Rayleigh particle, responds only to the statistical $\bi{E}$ field at its position, i.e. the coherence matrix.
It is therefore possible that classic problems such as atmospheric radiative transfer \cite{chandrasekhar:radiative} may be analysed using the $3\times 3$ coherence matrix.

A natural experimental situation in which the nonparaxial coherence matrix is  relevant is in the optical near field, for which measurements of the 3-dimensional field are possible \cite{ndsh:measuring} (of course the theory is not restricted to optical frequencies). 
The geometric interpretation should provide insight into the ensemble of polarization ellipses which gives rise to a measured $3\times 3$ coherence matrix.

\section*{Acknowledgements}

I am grateful to Michael Berry and John Hannay for useful discussions, and Girish Agarwal for pointing out to me the connection with density matrices in atomic physics. 
This work was supported by the Leverhulme Trust.


\end{document}